\documentclass[conference]{IEEEtran}
\usepackage{graphicx}
\usepackage{latexsym}
\usepackage{amsfonts}
\usepackage{amssymb}
\usepackage{amsbsy}
\usepackage{color}
\usepackage{amsmath}
\usepackage{multicol}
\usepackage{float}
\usepackage{subfigure}
\usepackage{algorithm}
\usepackage{algorithmic}
\usepackage[dvips]{epsfig}

\begin{document}
%
\title{ \LARGE Cognitive Random Access for Internet-of-Things Networks}


\author{\IEEEauthorblockN{Hyesung Kim$^\dagger$, Seung-Woo Ko$^*$, and Seong-Lyun Kim$^\dagger$}
\IEEEauthorblockA{
$^\dagger$School of Electrical and Electronic Engineering, Yonsei University,
Seoul, Korea\\
$^*$Department of Electrical and Electronic Engineering, The University of Hong Kong\\
Email: $^\dagger$\{hskim, slkim\}@ramo.yonsei.ac.kr, $^*$swko@eee.hku.hk}
}

\maketitle

\begin{abstract}

This paper focuses on cognitive radio (CR) internet-of-things (IoT) networks where spectrum sensors are deployed for IoT CR devices, which do not have enough
hardware capability to identify an unoccupied spectrum by themselves. 
In this sensor-enabled IoT CR network, the CR devices and the sensors are separated. It induces that spectrum occupancies at locations of CR devices and sensors could be different. 
To handle this difference, we investigate a conditional interference distribution (CID) at the CR device for a given measured interference at the sensor. 
We can observe a spatial correlation of the aggregate interference distribution through the CID.
Reflecting the CID, we devise a cognitive random access scheme which adaptively adjusts transmission probability  with respect to
the interference measurement of the sensor. 
Our scheme improves area spectral efficiency (ASE) compared to a conventional ALOHA and an adaptive transmission scheme which attempts to send data when the sensor measurement is lower than an interference threshold.


\end{abstract}


%
\IEEEpeerreviewmaketitle

\section{Introduction}

The increase of wireless internet-of-things (IoT) requires a large volume of vacant frequency bands that the current dedicated spectrum policy cannot cope with. To handle the spectrum shortage, the devices need to detect and access an unoccupied spectrum in an opportunistic manner \cite{CR_1}-\cite{C IoT}. However, most wireless IoT devices are hard to perform precise spectrum sensing by themselves due to their limited hardware capability and less cost  \cite{W IoT}. Spectrum sensors are essential as a part of the infrastructure for the sole purpose of interference monitoring \cite{infra}.


The locational difference of an IoT device and the corresponding sensor causes an observation error of spectrum occupancy. A mathematical model reflecting this spatial relationship is thus required. To this end, we derive a conditional interference distribution (CID) at an IoT device for a given measured interference from the sensor using stochastic geometry (SG). 
We find that its shape is skewed to the left of the sensor measurement and it has a long tail to the right side. This asymmetric tendency becomes intensified with increasing the measured interference level at the sensor. In other words, the IoT devices may experience less interference than the measured value with high probability. From the perspective of an opportunistic spectrum access, the IoT device would has more chances to exploit the band while guaranteeing the quality of services (QoS) of primary communications.  It is worth mentioning that the existing interference distributions based on SG \cite{SG_stoyan}--\cite{SG_Andrew} cannot explain the above asymmetric spatial correlation between the sensors and the IoT devices, which is a key to design cognitive radio IoT networks.

%

We propose a novel cognitive random access algorithm to adjust its transmission probability in a distributed manner according to the measured interference. Based on the CID, the proposed algorithm improves an area spectral efficiency (ASE) in return while satisfying the requirement of primary users. Analytic and numerical results show that our algorithm outperforms a conventional ALOHA \cite{p_s ALOHA} and a threshold based random access protocol with hard decision, where an IoT device can access the medium only when the measured sensor value is lower than the predetermined threshold.


\begin{figure}
\centering
\includegraphics[angle=0, height=0.18\textwidth]{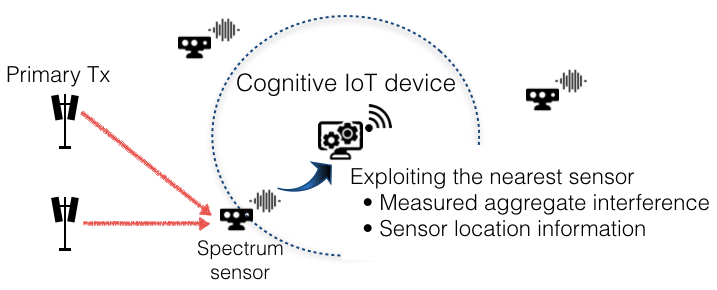}   
\caption{Cognitive IoT networks with spectrum sensors}\label{scenario}
\end{figure}

\section{Interference Distribution Conditioning on Sensor measurement}

\subsection{System Model}

In cognitive IoT networks with spectrum sensors, primary and IoT services try to access a shared spectrum band. The primary transmitter (PT) has a license to access the spectrum. 
The IoT transmitter, denoted as secondary transmitter (ST),  may acquire opportunistic access to the spectrum by exploiting the sensor measurement as shown in Fig. \ref{scenario}. 

\begin{figure}
\centering
\includegraphics[angle=0, height=0.18\textwidth]{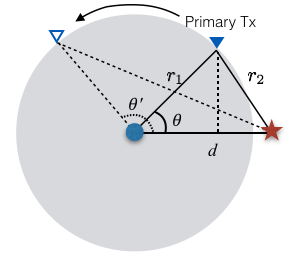}   
\caption{Schematic diagrams including a spectrum sensor (blue circle), its nearest PT (blue triangle), and the adjacent ST (red pentagram). }\label{geo}
\end{figure}

Consider a pair of ST and its adjacent sensor, which is located at the center as shown in Fig. \ref{geo}. 
The distance between the sensor and the ST is $d$. They are surrounded by PTs, whose locations follow a Poisson point process (PPP) $\Phi_p=\{x_1,x_2,...\} \in \mathbb{R}^2$ of density $\lambda_p$. 
The locations of the STs follow another independent PPP $\Phi_s=\{y_1,y_2,...\} \in \mathbb{R}^2$ of density $\lambda_s$. 

PTs and STs in the network use transmission powers $P_p$ and $P_s$, respectively.  We consider distance-dependent path loss, where $l(x,y)=\text{min}\{1,||x-y||^{-\alpha}\}$. Parameter $\alpha >2$ is a path-loss exponent. Fading is modeled as an  independent and identical random variable $h$.
Transmissions made by PTs impose an aggregate interference $I_{\Phi_p}$ at the sensor as follows:
\begin{equation}
I_{\Phi_p}=\sum_{z \in \Phi_p} P_p h_zl(z,o). 
\end{equation}
In this paper, we assume that spectrum sensors use energy detection \cite{ED} and measure the aggregate interference $I_{\Phi_p}$ without an error.
 
\subsection{Conditional Interference Distribution}

We focus on an interference distribution at the point  that is at a distance of $d$ from the sensor, when  the sensor measurement $I_m$ is given. 
We specify the conditional interference distribution (CID) function as
\begin{equation}
f_{I;m}(x)= \textbf{Pr}\{I=x|I_m=m\}.\label{CID_def}
\end{equation}
To derive the CID \eqref{CID_def}, we consider geometric situations, where PTs impose an aggregate interference to a sensor and its adjacent ST as depicted in Fig. \ref{geo}.
Let  $r_1$ denote the distance between the sensor and its nearest PT. The variable $r_2$ is the distance between the PT and the corresponding ST. 
Two lines from the sensor to the PT and the ST form an angle $\theta$ in radian unit. 

\vskip 5pt \noindent {\bf Proposition 1.} {\it For measured interference $I_m=m$, the CID function $f_{I;m}(x)$ is given by}
\setlength\arraycolsep{-1pt}
\begin{align}
&f_{I;m}(x)=\frac{\frac{P_p}{\pi d \hat{r}_1 \alpha  (x-T(\hat{r}_1,\alpha,\lambda_p))^2}\left(\frac{P_p}{x-T(\hat{r}_1,\alpha,\lambda_p)}\right)^{\frac{2}{\alpha}-1}}{
\sqrt{1-\frac{\left(\hat{r}_1^2+d^2-\left( \frac{P_p}{x-T(\hat{r}_1,\alpha,\lambda_p)}\right)^{\frac{2}{\alpha}}\right)^2}{4d^2\hat{r}_1^2} }},\label{CCID}
\end{align}
\noindent where $T(\hat{r}_1,\alpha,\lambda_p)= 2P\pi \lambda_p\frac{\hat{r}_1^{2-\alpha}}{\alpha-2} $, and $\hat{r}_1$ is a solution of the following equation: 
$\hat{r}^{\alpha}-\frac{2P_p\pi\lambda_p\hat{r}^2}{m(\alpha-2)}-\frac{P_p}{m}=0$.



%
\vskip 3pt \noindent{\it Proof:} Appendix.
\vskip 3pt
In Proposition 1, we consider that the nearest PT has a dominant effect on the CID, and approximate the distribution in terms of $\hat{r}_1$ denoting the estimated distance  between the sensor and its nearest PT. 
When the pathloss exponent $\alpha$ is 4, then the estimated distance $\hat{r}_1$ is equal to $\{\! (P_p\pi\lambda_p+\sqrt{(P_p\pi\lambda_p)^2+4m P_p})/2m\!\}^{0.5}$.

\begin{figure}
\centering
\includegraphics[angle=0, height=0.25\textwidth]{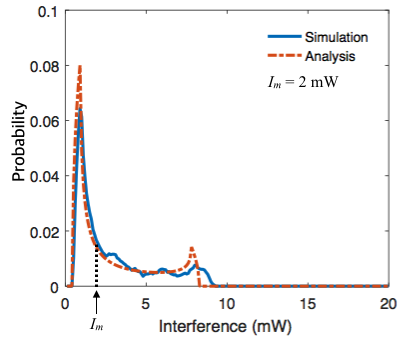}   
\caption{The conditional interference distributions  when $I_m=2$mW, $\alpha=4$, $d=1$, and $\lambda_p=0.003$, or 3000 PTs/km$^2$.}\label{simul_cid1}
\end{figure}

\begin{figure}
\centering
\includegraphics[angle=0, height=0.25\textwidth]{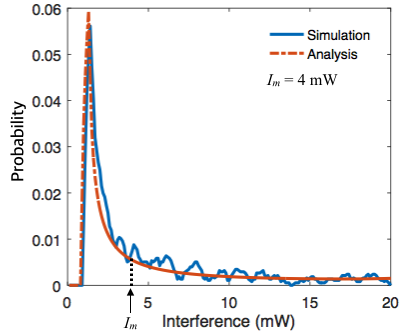}   
\caption{The conditional interference distributions when $I_m=4$mW, $\alpha=4$, $d=1$, and $\lambda_p=0.003$, or 3000 PTs/km$^2$.}\label{simul_cid2}
\end{figure}

Fig. \ref{simul_cid1} and Fig. \ref{simul_cid2} show two CIDs \eqref{CID_def} with respect to the different sensor measurement $I_m$.
The approximation of the CID \eqref{CCID} is tight when the PT density $\lambda_p$ is up to 0.003, or 3000 PTs/km$^2.$
It implies that the approximation would be effective to the case that the primary network is cellular network.

The shape of CID is  skewed to the left of the sensor measurement and characterized by having a long tail. 
Both of variance and skewness of the CID increase with  $I_m.$ 
It implies that the actual received interference at ST $i$ may be lower than the sensor measurement $I_m$ with considerable probability. In other words, there would be more transmission opportunities for STs. 
This phenomenon gives us an insight that STs can access the medium more aggressively while not degrading the primary communication qualities.

\section{Cognitive Random Access based on Conditional Interference Distribution }

\begin{figure}
\centering
\includegraphics[angle=0, height=0.25\textwidth]{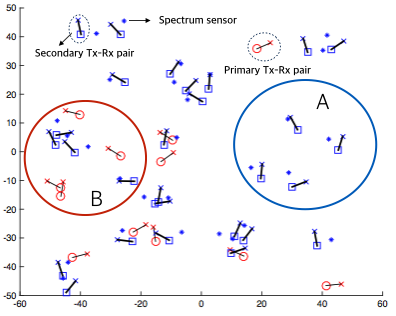}   
\caption{A snapshot of network topology. The PT density $\lambda_p$ is 0.001 and the ST density $\lambda_s$ is 0.002.
}\label{simul_topo}
\end{figure}

Fig. \ref{simul_topo} shows the snapshot of the network topology where PTs and STs are randomly located with respective densities of $\lambda_p=0.001$, and $\lambda_s=0.002$. Although the same density is applied to the network, we can observe the regional variance of the population, which makes different local interference conditions.  
The STs in subarea A are located in relatively sparse environment with low population of PTs, where the STs could access the channel without interruption. 
On the other hand, the STs in subarea B are located in relatively high region interference imposed by PTs, and the transmission attempt of STs in B would be obstructed.

To deal with these regional differences, we propose a cognitive random access by tuning each user's transmission probability based on the CID (3). 
It is worth noting that the CID gives us the probability that the aggregate interference at an ST is lower than an arbitrary threshold.
Unlike the conventional ALOHA, STs have the different transmission probabilities with respect to the interference measured by their adjacent sensors. 

We assume that the time is slotted and synchronized in the network. 
We focus on a snapshot of the communication process, where the network topology does not change during each time slot.
Each transmitter in the network always has enough data to transmit. 
Let us assume that the STs know the corresponding sensor locations. 
The time delay for which the ST receives interference measurement from its corresponding sensor is negligible. 

\vskip -9pt


\subsection{Improving Area Spetral Efficiency}

Our purpose is to improve the area spectral efficiency (ASE) $\eta$, the sum of data rates per unit bandwidth in the unit area \cite{ASE}, while protecting primary networks.  
Under a constraint of satisfying a required outage probability of primary transmissions, 
we determine transmission probabilities of STs $\textbf{p}=\{p_1,p_2,...,p_i,...\}$ in order to maximize $\eta$ as follows:
\begin{eqnarray} \label{eq:p1}
&&\hspace{-40pt}\textbf{(P1)}\quad\mathop{{\text{max}}}\limits_{\textbf{p}
} {\text{ }}\eta=\lambda_s \text{E}_i[ p_i] p_{s} \log(1+\beta)
 \\
&&\hspace{-40pt}\quad\quad\quad\text{subject to}\text{ }\quad\quad \text{Pr}\{SIR_p \leq \beta \}  \leq \tau, \label{eq:p1c1}\\
&&\hspace{-40pt}\text{ }\quad\quad\quad\quad\quad\qquad\qquad 0\leq p_i \leq1  \quad\forall i, \label{eq:p1c2}
\end{eqnarray}
\noindent where the notation $\beta$ denotes a target SIR threshold. The notation $p_s$ denotes a transmission success probability of secondary user.  We neglect noise in our analytical calculations. In the perspective of a typical ST, interferer density is given by $\lambda_p+\lambda_s\text{E}[p_i]$. From \cite{p_s ALOHA2}, the probability $p_s$ can be calculated as follows:
\begin{equation}
p_s=\text{Pr}[SIR>\beta]=e^{-(\lambda_p+\lambda_s\text{E}_i[ p_i]) r_s^2 \left(\frac{P_p\beta}{P_s}\right)^{2/\alpha}C(\alpha)},
\end{equation}
\noindent where $C(\alpha)\!=\!\frac{2\pi}{\alpha}\Gamma(\frac{2}{\alpha})\Gamma(1-\frac{2}{\alpha})$ and $\Gamma(z)=\int_0^\infty t^{z-1}e^{-t}dt$. The parameter $r_s$ is a distance between secondary transmitter and receiver. Here, we assume that the distance $r_s$ is same for all STs and their corresponding receivers. 
The constraint \eqref{eq:p1c1} assures that the outage probability of the primary communications cannot exceed the target value. 

\vskip 2pt \noindent {\bf Proposition 2.} {\it The optimal transmission probabilities} $\textbf{p}^*=\{p^*_1,p^*_2,...,p^*_i,...\}$ {\it should satisfy the following equation:}
\begin{equation}\label{Propo2}
\text{E}[p_i^*]\!=\!\max\!\left[0,\min\!\left\{\!1,\frac{1}{\lambda_s}\left(\frac{\ln\left(1/(1-\tau)\right)}{r_p^2\left(\frac{P_s\beta}{P_p}\right)^{2/\alpha}{\frac{2}{\alpha}}C(\alpha)}-\lambda_p\! \right)\!\right\}\!\right]\!. 
\end{equation}
 \noindent{\it
Proof}: From the outage probability in \cite{p_outage}, we can represent the constraint \eqref{eq:p1c1} as follows:
\begin{equation}
\textbf{Pr}\{SIR_p\leq \beta\} = 1-e^{-(\lambda_p+\lambda_s \text{E}[p_i])r_p^2 \left(\frac{P_s\beta}{P_p}\right)^{2/\alpha}C(\alpha)},
\end{equation}
\vskip-10pt
\begin{equation}\label{ineq}
 \text{E}[p_i]\leq\frac{1}{\lambda_s}\left\{\frac{\ln\left(1/(1-\tau)\right)}{r_p^2 \left(\frac{P_s\beta}{P_p}\right)^{2/\alpha}C(\alpha)}-\lambda_p \right\}.
\end{equation}
We can rewrite the expectation term $\text{E}[p_i]$ of \eqref{ineq} as $\frac{1}{N}\sum_i^Np_i$ without loss of generality, where $N$ is the number of STs.  
Then, we can relax constraint \eqref{eq:p1c1} and obtain the following Lagrangian function:
\begin{align}
L(\textbf{p},\mu)&=\lambda_s \text{E}[ p_i] p_{s} \log(1+\beta) \nonumber\\
&+\mu \left \{\frac{N}{\lambda_s}\left(\frac{\ln\left(1/(1-\tau)\right)}{r_p^2 \beta^{2/\alpha}C(\alpha)}-\lambda_p \right) -\sum_i^Np_i \right\},
\end{align}
where $\mu$ is a nonnegative Lagrangian multiplier. The Karush-Kuhn-Tucker (KKT) conditions of the problem \textbf{P1} are necessary for optimality. The  KKT conditions  are given as follows:
\vskip -10pt
{\small
{\medmuskip=-0.5mu\thinmuskip=-0.5mu\thickmuskip=-0.5mu\begin {align}\label{kkt1}
 \frac{\partial L}{\partial p_i}\!&=\frac{\lambda_p \lambda_s\log(1+\beta)e^{-(\lambda_p+\lambda_s/N\sum_i^Np_i)r_s^2 \left(\frac{P_p\beta}{P_s}\right)^{2/\alpha}C(\alpha)}}{N} \nonumber\\
  &\quad\times\underbrace{\left(1-\frac{\lambda_s}{N}\sum_i^Np_i r_s^2\left(P_p\beta/P_s\right)^{2/\alpha}C(\alpha)\right)}_{(A)}-\mu \leq 0,
\end{align}}
\vskip -14pt
\begin {equation}\label{kkt2}
 \frac{\partial L}{\partial \mu}= \frac{N}{\lambda_s}\left(\frac{\ln\left(1/(1-\tau)\right)}{r_p^2\left(\frac{P_s\beta}{P_p}\right)^{2/\alpha}C(\alpha)}-\lambda_p \right) -\sum_i^Np_i \leq 0,
\end{equation}
\vskip -10pt
{\medmuskip=-0.5mu\thinmuskip=-0.5mu\thickmuskip=-0.5mu\begin{align}
\!p_i&\Bigg\{\underbrace{\frac{\lambda_p \lambda_s\log(1+\beta)e^{-(\lambda_p+\lambda_s/N\sum_i^Np_i)r_s^2 \left(\frac{P_p\beta}{P_s}\right)^{2/\alpha}C(\alpha)}}{N}}_{(B)}\nonumber \\ 
 \!\!&\times\left(1-\frac{\lambda_s}{N}\sum_i^Np_i r_s^2\left(P_p\beta/P_s\right)^{2/\alpha}C(\alpha)\right)-\mu\Bigg\}=0, \label{kkt3}
\end{align}}
\begin{equation}\label{kkt4}
\mu \left\{\frac{N}{\lambda_s}\left(\frac{\ln\left(1/(1-\tau)\right)}{r_p^2\left(P_s\beta/P_p\right)^{2/\alpha}C(\alpha)}-\lambda_p \right) -\sum_i^Np_i \right\}=0,
\end{equation}
\begin{equation}\label{kkt5}
0\leq p_i \leq1  \quad\forall i.
\end{equation}}
The variable $\mu$ should be positive. It can be proved as follows.  When $\mu=0$, the variable $p_i$ should be zero for all $i$ since the term ($B$) is always positive. Then, the term ($A$) always has  a positive value, and makes the partial derivative $\partial L/\partial p_i$ positive. It does not satisfy the condtion \eqref{kkt1}. Therefore, 
\vskip -2pt {\small
\begin{equation}
\sum_i^Np_i =\frac{N}{\lambda_s}\left(\frac{\ln\left(1/(1-\tau)\right)}{r_p^2\left(\frac{P_s\beta}{P_p}\right)^{\frac{2}{\alpha}}C(\alpha)}-\lambda_p \right), 
\end{equation}}
and the variable $\mu$ is positive and equal to the term ($B$).
\hfill $\blacksquare$

\begin{algorithm}
\centering
\caption{Cognitive Random Access}\label{euclid}
\begin{algorithmic}[1]
\REQUIRE $\lambda_p$, $\lambda_s$, and $I_{th}$
\STATE $\textbf {Spectrum sensor}$: 
\STATE Measure the aggregate interference $m_i$
\STATE $\textbf {Secondary transmitter}$: 
\STATE Get $m_i$ from the spectrum sensor
\STATE Compute a probability $F_{I;m_i}(I_{th})$ for $m_i$ using \eqref{CCID}
\STATE $w_i \Leftarrow F_{I;m_i}(I_{th})$
\STATE Caculate an expectation $E[p_i^*]$ \\
\vskip 1pt
$E[p_i^*]\Leftarrow \max\!\left[0,\min\!\left\{\!1,\frac{1}{\lambda_s}\left(\frac{\ln\left(1/(1-\tau)\right)}{r_p^2\left(\frac{P_s\beta}{P_p}\right)^{2/\alpha}{\frac{2}{\alpha}}C(\alpha)}-\lambda_p\! \right)\!\right\}\!\right]\!$ 
\STATE $p_i\Leftarrow \min\left[1, \frac{w_i}{E[w_i]}E[p_i^*]\right]$
\STATE Go line 2
\end{algorithmic}
\end{algorithm}

The Proposition 2 determines only the expectation $E[p_i^*]$ of the optimal transmission probability $p_i^*$. Therefore, we need to find the optimal $p_i^*$ for each ST $i$. 
Combining \eqref{CCID} in Proposition 1 and \eqref{Propo2} in Proposition 2, we devise an algorithm that determines the suboptimal transmission probability $\hat{p}_i$ for each ST $i$ as follows.

We propose a simple algorithm to find $\hat{p}_i$ that satisfies the necessary condition \eqref{Propo2}. Let $w_i$ be a probability weight factor for ST $i$. The transmission probability $\hat{p}_i$ is determined as follows:
\begin{equation}\label{p^*}
\hat{p}_i=\min\left[1, \frac{w_i}{E[w_i]}E[p_i^*]\right],
\end{equation}
Here, the cumulative CID $F_{I;m_i}$ determines the factor $w_i$ in the following manner: $w_i=F_{I;m_i}(I_{th})$,
where $m_i$ is a measured interference by the sensor adjacent to the ST $i$, and $I_{th}$ is a given interference threshold.  
The weight $w_i$ means the probability that the ST $i$ receives an aggregate interference lower than the threshold $I_{th}$. 
In \eqref{p^*}, the normalized probability $w_i/E[w_i]$ adjusts a chance of spectrum access.
For example, as shown in Fig \ref{simul_topo}, the STs in sparse environment like area A may have a high weight $w_i$.    
The whole process of the proposed cognitive random access is described in Algorithm 1.

\begin{figure}
\centering
\includegraphics[angle=0, height=0.26\textwidth]{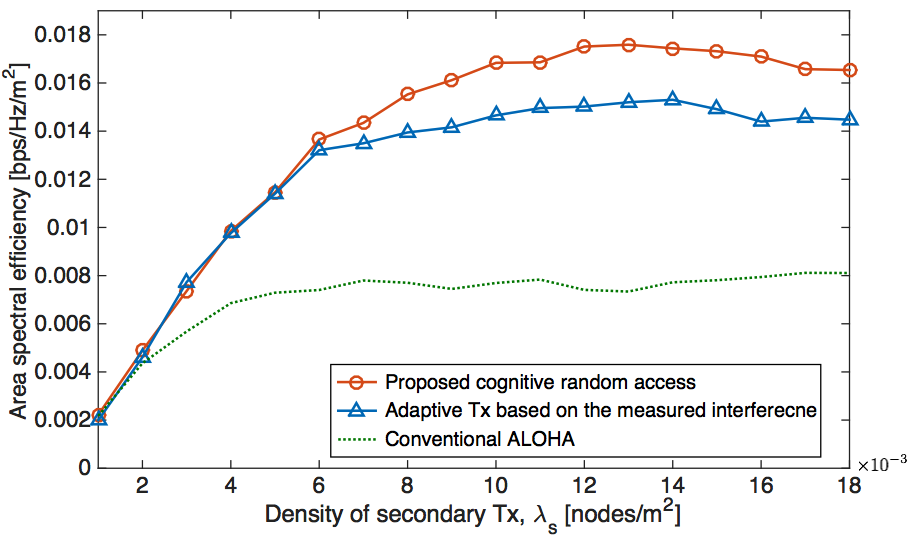}   
\caption{ASE as a function of secondary transmitter density ($\lambda_p=0.001, \beta=  3 \text{ dB}, \tau=0.05,d=1$).  }\label{fig ASE_s}
\end{figure}

\begin{figure}
\centering
\includegraphics[angle=0, height=0.27\textwidth]{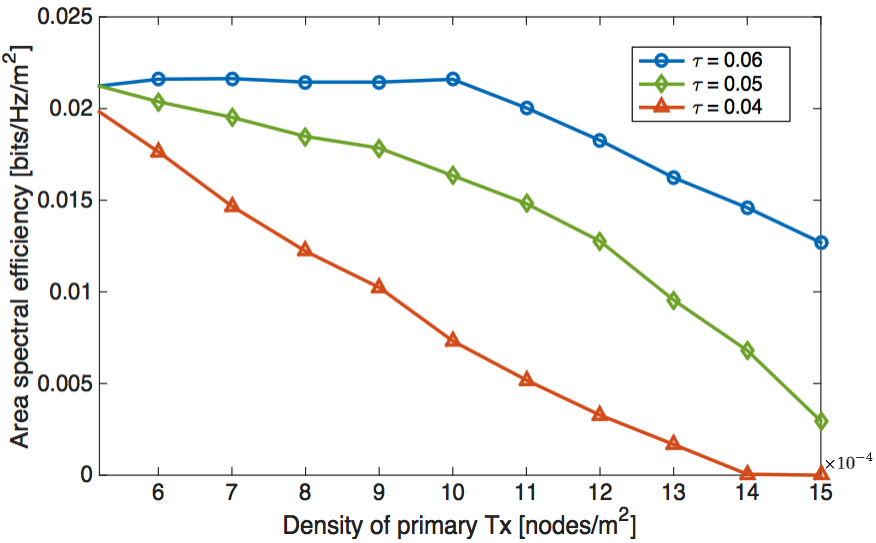}   
\caption{ASE as a function of primary transmitter density with various $\tau$ ($\lambda_s=0.01, \beta=3\text{ dB},d=1)$.  }\label{fig ASE_p}
\end{figure}

\subsection {Performance Evaluation}

We evaluate  the proposed random access scheme through
1,000,000 simulations. At every simulation, PTs and STs are independent and identically distributed according to a homogeneous PPP with intensity $\lambda_p$ and $\lambda_s$, respectively, in a 100 m $\times$ 100 m area. Spectrum sensor is located at a distance of $d$ from the corresponding ST.
The communication distances of primary and secondary pairs is 3 m. The distance between the ST and its adjacent sensor is set to 1 m. 
The interference threshold $I_{th}$ is set to 2 dBm.  The primary and secondary transmission powers are 23 dBm and 5 dBm, respectively. We set the noise power as -70 dBm and consider Rayleigh fading.

Fig. \ref{fig ASE_s} shows ASE performance as a function of the ST density $\lambda_s$.
The proposed scheme surpasses the conventional ALOHA scheme. Also, we conducted the comparison with adaptive transmission that deterministically attempts to send data 
when the measurement of an adjacent sensor is lower than an interference threshold.
The proposed scheme shows better ASE performance than the other schemes for high density of ST $\lambda_s$. 
This ASE difference comes from the probabilistic transmission based on the CID \eqref{CCID}, which gives us a probability that actual received interference at ST may be lower than the measurement. 
With this information, the ST attempts to access the spectrum more aggressively, producing higher ASE. Also, we observe ASE performance of the proposed scheme with respect to the PT density $\lambda_p$ as shown in Fig. \ref{fig ASE_p}. When the primary outage probability constraint $\tau$ is small, ASE is more sensitive to $\lambda_p$.



\section{Conclusion}

This paper focuses on cognitive radio (CR) based IoT networks where multiple sensors are deployed to monitor interference temperature in the area.
An inherent characteristic of this CR network induces the different spectrum occupancy at the locations of the CR IoT devices and the sensors. 
To compensate this difference, we derive a conditional interference distribution (CID) at the CR device for a given measured interference at the sensor. 
We find that the shape of the CID is a left-skewed distribution. This statistic property implies that an actual received interference at CR device may be 
lower than the sensor measurement with a considerable probability. Reflecting this phenomenon, we devise a cognitive random access scheme 
which adaptively adjusts transmission probability  with respect to the CID.
Our scheme improves area spectral efficiency compared to conventional ALOHA and threshold based random access protocol, where an IoT device can access the medium only when a measured sensor value is lower than the predetermined threshold.

\section{Appendix: Proof of Proposition 1}

The measured aggregate interference $I_m$ at the sensor can be decomposed as follows:
\begin{equation}
I_m=I_{x_1}+ I_{\sum_{\Phi_p\setminus \{x_1\}} },\label{decomp}
\end{equation} 
\noindent where $I_{x_1}=Pl(o,r_1)$, $x_1$ is the nearest PT from the sensor, and $r_1$ is the distance between the sensor and  PT $x_1$.
Since spectrum sensors measure interference for a enough time duration, the fading effect can be averaged out in the measurement. 
We assume that sensors are close to STs enough to consider PT $x_1$ as a common dominant interferer for both the sensor and  ST $i$.
Now then, we investigate how these two sets of interferers have an influence on the ST $i$.
First, PT $x_1$ imposes an interference $I_{x_1}$ to ST $i$ as $I_{x_1}=P({r_1^2+d^2-2r_1d\cos\theta})^{-\alpha/2}$.
For a fixed interference $Pr_1^{-\alpha}$, the angle $\theta$ determines interference strength $I_{x_1}$. 
As $\theta$ increases, the interference $I_{x_1}$  decreases by $\cos(\theta)$. Using this geometric property,
we can transit the cumulative CID function $F_{I;m}(x)=\textbf{Pr}\{I\leq x|I_m=m\}$ to the probability with respect to $\theta$ as follows:
\begin{align}
&\textbf{Pr}\{\theta \geq \theta_x|I_m=m\}=1-\frac{\theta_x}{\pi}\\
&=1-\pi\cos^{-1}\Bigg(\frac{r_1^2+d^2-\Big(\frac{P}{x-\sum_{\Phi_p\setminus \{x_1\}} Pl(o,r_i)}\Big)^2}{2r_1d}\Bigg). \label{Pr}
\end{align}
Here, we use an approximation that interference $I_{\sum_{\Phi_p\setminus \{x_1\}} }$ at ST $i$ from PTs in $\Phi_p\setminus \{x_1\}$ is equal to mean interference $E[I_{\Phi_p\setminus \{x_1\}}]=\lambda_p \int_0^{2\pi}\!\!\!\int_{r_1}^\infty Pl(o,r)  rdrd\phi=2P\pi \lambda_p\frac{r_1^{2-\alpha}}{\alpha-2} $.
Let $\hat{r}_1$ denote  the estimated distance from the nearest PT. For a given $I_m$ equal to $m$ in \eqref{decomp}, we can find the estimated distance 
$\hat{r}_1$ which is the solution of the following equation: $m=P\hat{r}_1^{-\alpha}+2P\pi \lambda_p\frac{\hat{r}_1^{2-\alpha}}{\alpha-2}$. 
Now, then we can specify the cumulative CID $F_{I;m}(x)$, of which derivative is the CID function $f_{I;m}(x)$ \eqref{CCID}.  
{\small{
\begin{align}
&\frac{d}{dx}\left\{1-\pi\cos^{-1}\left(\frac{r_1^2+d^2-\left(\frac{P}{x-2P\pi \lambda_pr_1^{2-\alpha}{/(\alpha-2)}}\right)^2}{2r_1d}\right)\right\} \nonumber \\
&= \quad \frac{\frac{P_p}{\pi d \hat{r}_1 \alpha  (x-T(\hat{r}_1,\alpha,\lambda_p))^2}\left(\frac{P_p}{x-T(\hat{r}_1,\alpha,\lambda_p)}\right)^{\frac{2}{\alpha}-1}}{
\sqrt{1-\frac{\left(\hat{r}_1^2+d^2-\left( \frac{P_p}{x-T(\hat{r}_1,\alpha,\lambda_p)}\right)^{\frac{2}{\alpha}}\right)^2}{4d^2\hat{r}_1^2} }},
\end{align}}}
where $T(\hat{r}_1,\alpha,\lambda_p)=2P\pi \lambda_p\frac{\hat{r}_1^{2-\alpha}}{\alpha-2}$. \hfill $\blacksquare$


%
%

\section*{Acknowledgement}

This research was supported by the Institute for Information \& communications Technology Promotion (IITP) grant funded by the Korea government (MSIP) (No. 2015-0-00294, Spectrum Sensing and Future Radio Communication Platforms and No. 2016-0-00208, High Accurate Positioning Enabled MIMO Transmission and Network Technologies for Next 5G-V2X (vehicle-to-everything) Services).



%

\end{document}